\documentclass[12pt]{iopart}
\usepackage[pdftex]{graphicx,color}
\usepackage{url}
\begin{document}

\title[Granular deformation by centrifuge]{Measurement of surface deformation and cohesion of a granular pile under the effect of centrifugal force}

\author{Terunori Irie$^1$, Ryusei Yamaguchi$^2$, Sei-ichiro Watanabe$^1$,\& Hiroaki Katsuragi$^3$}

\address{$^1$Department of Earth and Environmental Sciences, Nagoya University, Furocho, Chikusa, Nagoya 464-8601, Japan \\
$^2$Technical Center, Nagoya University, Furocho, Chikusa, Nagoya 464-8601, Japan \\
$^3$ Department of Earth and Space Science, Osaka University, 1-1 Machikaneyama, Toyonaka 560-0043, Japan
}
%\ead{submissions@iop.org}
%\vspace{10pt}
%\begin{indented}
%\item[]August 2017
%\end{indented}

\begin{abstract}
An experimental apparatus measuring free-surface deformation of a centrifuged granular pile is developed. By horizontally rotating a quasi two-dimensional granular pile whose apex is located at the vertical rotation axis, the resultant force of gravity and centrifuge yields the deformation of the granular pile. In this setup, centrifugal force depends on distance from the rotation axis whilst gravitational force is constant everywhere. Therefore, free-surface deformation by various centrifuge degrees can be systematically examined using this apparatus. In the system, a small unit consisting of a camera and computer is rotated with the granular sample to record the rotation-induced deformation. To evaluate the validity of the system, deformation of a rotated water surface is first measured and analyzed. The obtained data are properly explained by the theoretical parabolas without any fitting parameter. Next, we measure the deformation of non-cohesive and cohesive granular piles using the developed apparatus. Both granular samples show the significant deformation of granular pile and finally develop steep granular slopes on the side walls. However, details of the deformation processes depend on the cohesion strength. To quantitatively characterize the difference, the effective strength by cohesion and granular local-slope variations are analyzed based on the experimental results.
\end{abstract}

%\maketitle

\section{Introduction}
\label{sec:introduction}
Rheological characterization of granular matter has been a crucial issue to understand its physical behaviors under various conditions~\cite{Andreotti:2013}. In many scientific and industrial fields, scientists and engineers must properly understand and/or handle various kinds of granular materials. For example, flowability of granular matter is an important key to effectively handle powdery and granular materials in chemical engineering processes~\cite{Rao:2008}. In planetary science, researchers have realized that a lot of small bodies in the solar system can be regarded as granular matter, so-called rubble-pile asteroids~\cite{Walsh:2018}. Specifically, a recent exploratory mission has examined the shape and structure of an asteroid Ryugu as a demonstrative example of rubble-pile asteroids~\cite{Watanabe:2019,Okada:2020}. Top-shape form of the asteroid Ryugu is presumably determined by the resultant balance of the spin-induced centrifugal force and the self-gravitational force. Although this idea is qualitatively reasonable, much more details on the granular behaviors under the effects of centrifugal and gravitational forces have to be revealed to quantitatively understand the asteroids' top shapes. 

Regarding gravity effect, granular behaviors under the microgravity conditions have been studied extensively in various works. As an essential fundamental feature, gravity dependence of granular angle of repose has been studied. Using rotating drum experiments, angle of repose under various gravitational conditions has been measured~\cite{Klein:1990,Arndt:2006,Cosby:2009,Kleinhans:2011}. Other experiments such as avalanching and pushing have also been performed to characterize gravity dependence of angle of repose~\cite{Hofmeister:2009,Blum:2010,Marshall:2018,Chen:2019}. Most of these studies have reported that angle of repose is almost independent of or weakly dependent of gravity. However, the results are not fully consistent with each other. While the definition of angle of repose is simple, its actual value depends on the measuring methods. Moreover, there are at least two types of angles characterizing the stability of granular slopes: {\it{static}} and {\it{dynamic}} angles that could be affected by grain size~\cite{Nagel:1992}. The static angle (starting angle) indicates the maximumly stable angle of a granular slope and the dynamic one (angle of repose) represents the settling angle of a granular slope after avalanching. However, any physical quantity appropriately characterizing the difference of these two angles has not been established. To evaluate the precise gravity dependence of granular angles, a novel measuring method of granular-pile deformation is necessary. Centrifugal system is a possible mechanism to study the general body-force dependence of granular angles.

In fact, centrifugal methods have been widely used for a number of granular-related experiments. Examples include the centrifuge-based dehydration~\cite{Bizard:2013}, dredge pumping~\cite{Hong:2016}, soil characterization~\cite{Gurung:1998}, silo discharging~\cite{Dorbolo:2013}, surface deformation~\cite{Cabrera:2017}, cohesion measurement~\cite{Nagaashi:2018,Nagaashi:2021}, and mechanical characterization~\cite{Castellanos:2007,SoriaHoyo:2008,Herminghaus:2013}. In general, centrifuge is useful to separate, transport and deform various materials. Thus, the fundamental understanding of the effect of centrifugal force on granular behaviors is also crucial to design efficient handling instruments in industry. All the above-mentioned instruments are basically motivated by the very specific applications to industrial or planetary phenomena. 

In this study, we develop a new apparatus which utilizes the centrifugal force to deform the granular pile. Using the developed experimental system, deformation of granular free surface caused by the centrifugal force is precisely acquired and analyzed. Particularly, difference between cohesive and non-cohesive grains is analyzed on the basis of experimental results. The developed system is similar to the apparatus reported in \cite{Castellanos:2007,SoriaHoyo:2008}. In these studies, flowability has been evaluated using a rotating two-dimensional granular layer. However, they have used an initially horizontal surface. Namely, deformation of the granular pile which initially has an angle of repose has not been investigated thus far. This initial condition (granular pile with angle of repose) is crucial to study the granular-related terrain dynamics. In the literature, relaxation of granular slope due to the vibration~\cite{Roering:2001a,Tsuji:2018}, small-scale impacts~\cite{Soderblom:1970,Omura:2020} have been studied to model the planetary terrain dynamics. To the best of our knowledge, however, relaxation of a granular pile due to the centrifuge has not been studied experimentally. Thus, we build a new experimental setup to reveal the mechanics of granular pile relaxation due to the competition between the centrifuge and gravity. 

In summary, we develop a novel experimental apparatus by which the free-surface deformation of granular piles driven by centrifugal force can be precisely measured. Using this system, we analyze the effect of cohesion on granular surface deformation. From the obtained data, fundamental aspects of a rotated granular pile (e.g., centrifuge-dependent local slope and effective cohesion strength) can be estimated. In this paper, technical features of the developed system and typical measured results (particularly the estimate of effective cohesive strength) are presented. 

\section{Experimental system}
\subsection{Experimental apparatus}
\label{sec:appratus}
\begin{figure}
\begin{center}
\includegraphics[width=.75\linewidth]{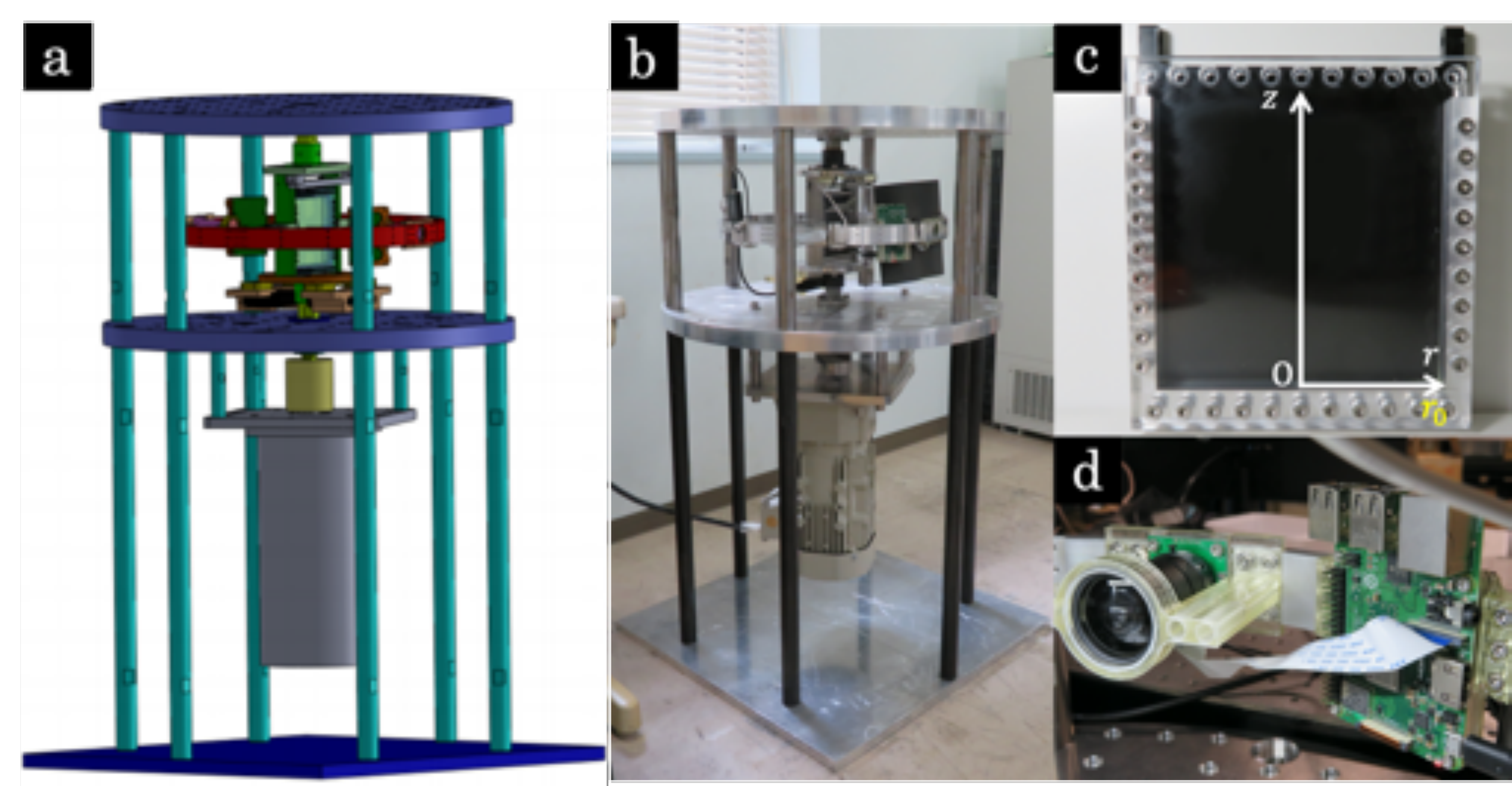}
\end{center}
\caption{Schematic drawing and pictures of the experimental apparatus. (a)~A design of the apparatus is presented. Most of the frames are made of stainless steel. The motor is mounted beneath the rotational cell to lower the center of mass of the system. The upper rotational part is covered during the experiment (not shown). Pictures of (b)~rotating system, (c)~(empty) cell, and (d)~camera-computer unit are shown in each panel. The coordinate system is also shown in (c).}
\label{fig:apparatus}
\end{figure}

The experimental apparatus developed in this study is shown in Fig.~\ref{fig:apparatus}. Figure~\ref{fig:apparatus}(a) shows a schematic design of the system. A quasi two-dimensional cell made of acrylic plates and aluminium bars with inner dimensions 100~mm$\times$100~mm$\times$10~mm is vertically held at the center of the apparatus. An actual image of the apparatus (without the cell) is shown in Fig.~\ref{fig:apparatus}(b). The cell is rotated around the vertical rotation axis by a motor (Sumitomo Heavy Industries, ZNFM1-1280-AP-3) mounted below the cell unit. The cell is carefully fixed so that the rotation axis corresponds to the center of the cell. Because the centrifugal force depends on distance from the rotation axis, horizontal distance from the rotation axis $r$ is an important parameter. In this experimental apparatus, the maximum value of $r$ is $r_0=50$~mm (distance from the cell center to the side wall). The vertically upward direction is defined as $z$ axis. The cell and coordinate system are shown in Fig.~\ref{fig:apparatus}(c). The front panel of the cell is made of a transparent acrylic plate to observe the deformation of the sample. In front of the cell, a small unit consisting of a camera (ArducamSKU:B0032) and a computer (RaspberryPi 3) is fixed to the rotating part (Fig.~\ref{fig:apparatus}(d)). The camera-computer unit is always facing the cell so that the camera-computer unit can acquire the cell images at arbitrary timing during the rotation (Fig.~\ref{fig:apparatus}(b)). The cell is illuminated by a LED light. A battery unit for the light and computer is also mounted on the rotation unit. The camera-computer unit is controlled by a desktop PC via a wireless connection. The rotation rate of the cell is controlled by an inverter (Sumitomo Heavy Industries, SF5202-1A5). During the rotation, a cover on the rotational part is attached for safety's sake whereas it is not shown in Fig.~\ref{fig:apparatus}.  

\subsection{Calibration}
\label{sec:calibration}
\begin{figure}
\begin{center}
\includegraphics[width=0.4\linewidth]{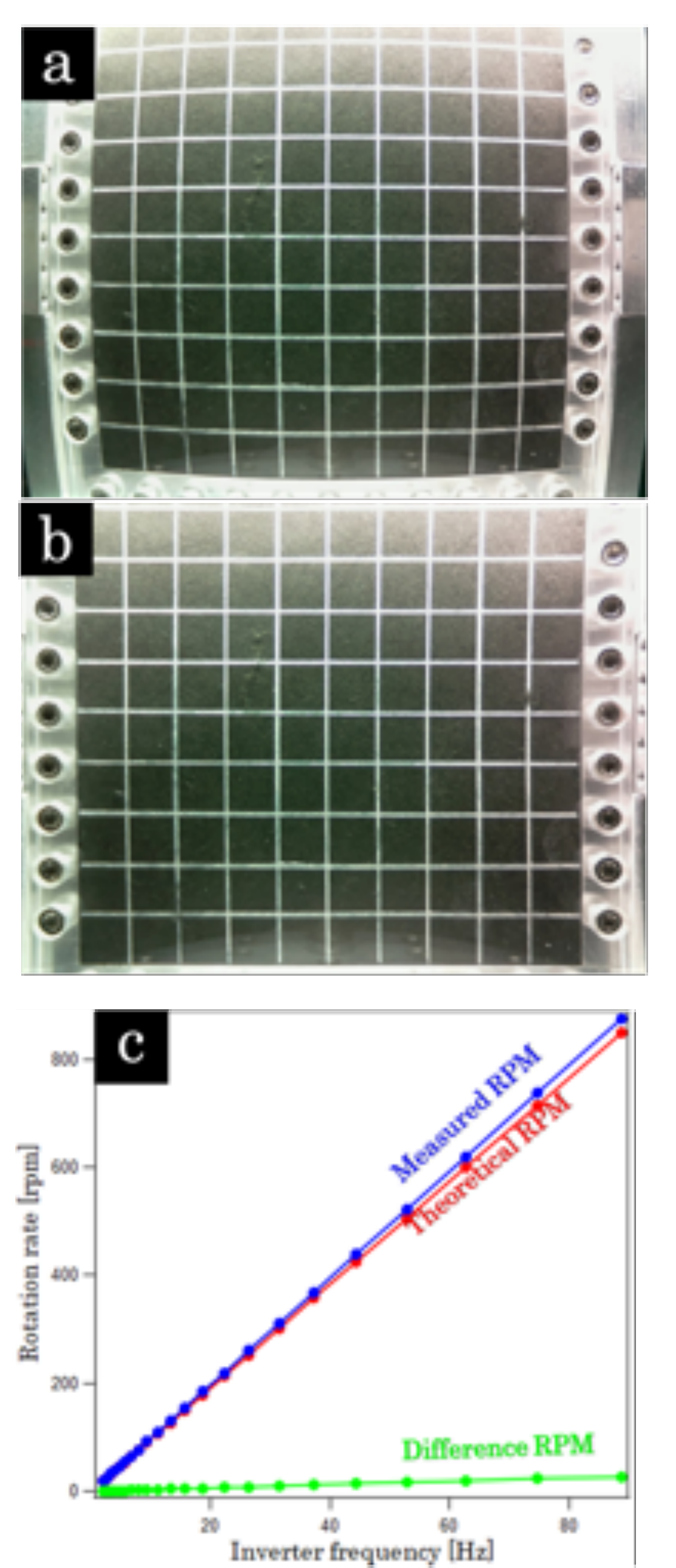}
\end{center}
\caption{Grid images of (a)~before and (b)~after the lens distortion correction. The square grid fixed on the surface of the empty cell is taken by the camera. The GIMP software is used for the correction. (c)~The relation between theoretical and measured rotation rates by varying the inverter's frequency is presented. Only a few percent error is detected.}
\label{fig:calibration}
\end{figure}

To obtain reliable data, appropriate calibrations are indispensable. First, we have to correct the distortion of the acquired image. The lens attached to the camera yields large image distortion as shown in Fig.~\ref{fig:calibration}(a). Although the metal frame and grid lines shown in the image should be straight, non-negligible distortion is clearly confirmed in Fig.~\ref{fig:calibration}(a). To correct the distortion, we use GIMP software. By using the lens-distortion function, the acquired images are corrected as displayed in Fig.~\ref{fig:calibration}(b). Because all the system configurations are fixed in all measurements, identical parameters for the lens-distortion correction (Main:-24.70, Edge:0.00, Zoom:0.00) are used to all images taken by this system. Spatial resolution of the acquired images is 0.048~mm/pixel and image size is 2592$\times$1944 pixels. 

The accuracy of the rotation rate is also evaluated. While the rotation rate is supposed to be controlled by the inverter's frequency, the accuracy of the rotation rate should be checked by an independent measurement. Therefore, actual rotation rates with various inverter's frequencies are measured by a tachometer (Line Seilki, TM-5010K). The relation between the inverter's frequency and the measured rotation rate is shown in Fig.~\ref{fig:calibration}(c). The actual rotation rate is slightly larger than the theoretically expected value. However, the error is at most a few percent level. Therefore, we can conclude the system works properly as expected. In the following, the corrected rotation rate based on this calibration measurement is used for the analysis.

To indicate the magnitude of centrifuge, we introduce a dimensionless parameter, 
\begin{equation}
  \Gamma = \frac{r_0 \omega^2}{g},
\label{eq:Gamma}
\end{equation}
which characterizes the strength of centrifugal force compared with gravitational force. Here, $\omega$ and $g=9.8$~m/s$^2$ are the rotational angular speed and gravitational acceleration, respectively. 

\subsection{Water surface deformation}
\label{sec:water_deform}
\begin{figure}
\begin{center}
\includegraphics[width=1.0\linewidth]{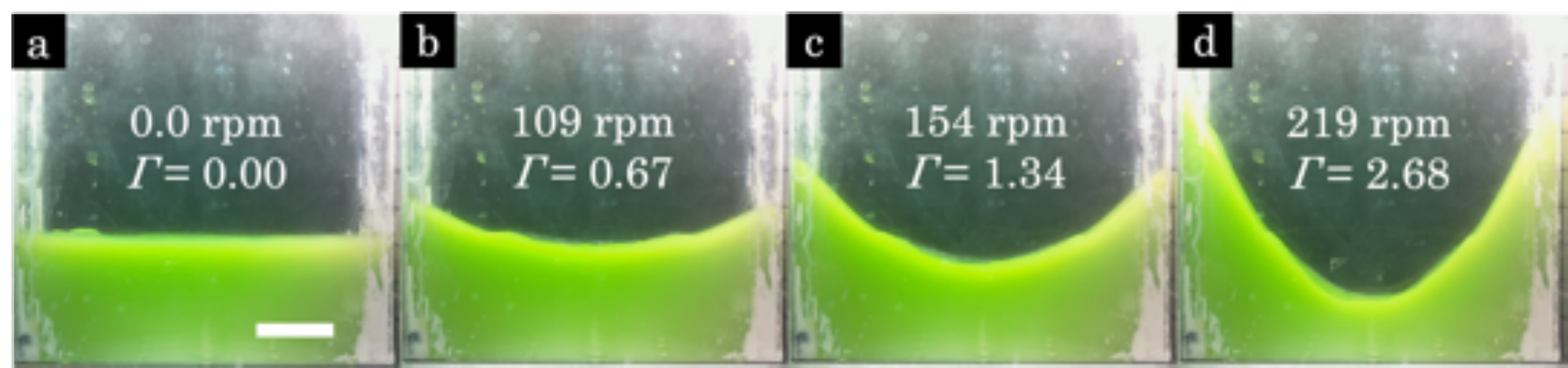}
\end{center}
\caption{Surface deformations of the rotated color water with (a)~$\Gamma=0$, (b)~$\Gamma=0.67$, (c)~$\Gamma=1.34$, and (d)~$\Gamma=2.68$ are shown. The scale bar indicates 20~mm. Parabola shapes can be confirmed in finite $\Gamma$ cases. }
\label{fig:water_raw}
\end{figure}

To evaluate the ability of the developed system, we first measure the deformation of water surface. Colored water is poured in the cell. Then, the cell is gradually rotated. In Fig.~\ref{fig:water_raw}, images of the rotated water surface at $\Gamma=0$, $0.67$, $1.34$, and $2.68$ are presented. Due to the centrifugal force, water surface is gradually deformed. The form of a rotated water surface can be computed by considering the force balance as, 
\begin{equation}
\frac{z}{r_0} = \frac{\Gamma}{2} \left( \frac{r}{r_0} \right)^2 + \frac{z_c}{r_0},
\label{eq:water}
\end{equation} 
where $z_c$ is an integral constant which can be determined by other conditions such as volume conservation. Here, we neglect the effect of $z_c$, mainly owing to the technical limitation of the apparatus. In this system, small but finite amount of water leakage from the sample cell is observed. Because the main aim of this apparatus is the measurement of granular surface deformation, the cell is not fully water tight. However, this has no impact on the local slope measurement. Thus, in this experiment, we focus on the parabola shape evaluation by neglecting $z_c$ term in Eq.~(\ref{eq:water}).

\begin{figure}
\begin{center}
\includegraphics[width=0.5\linewidth]{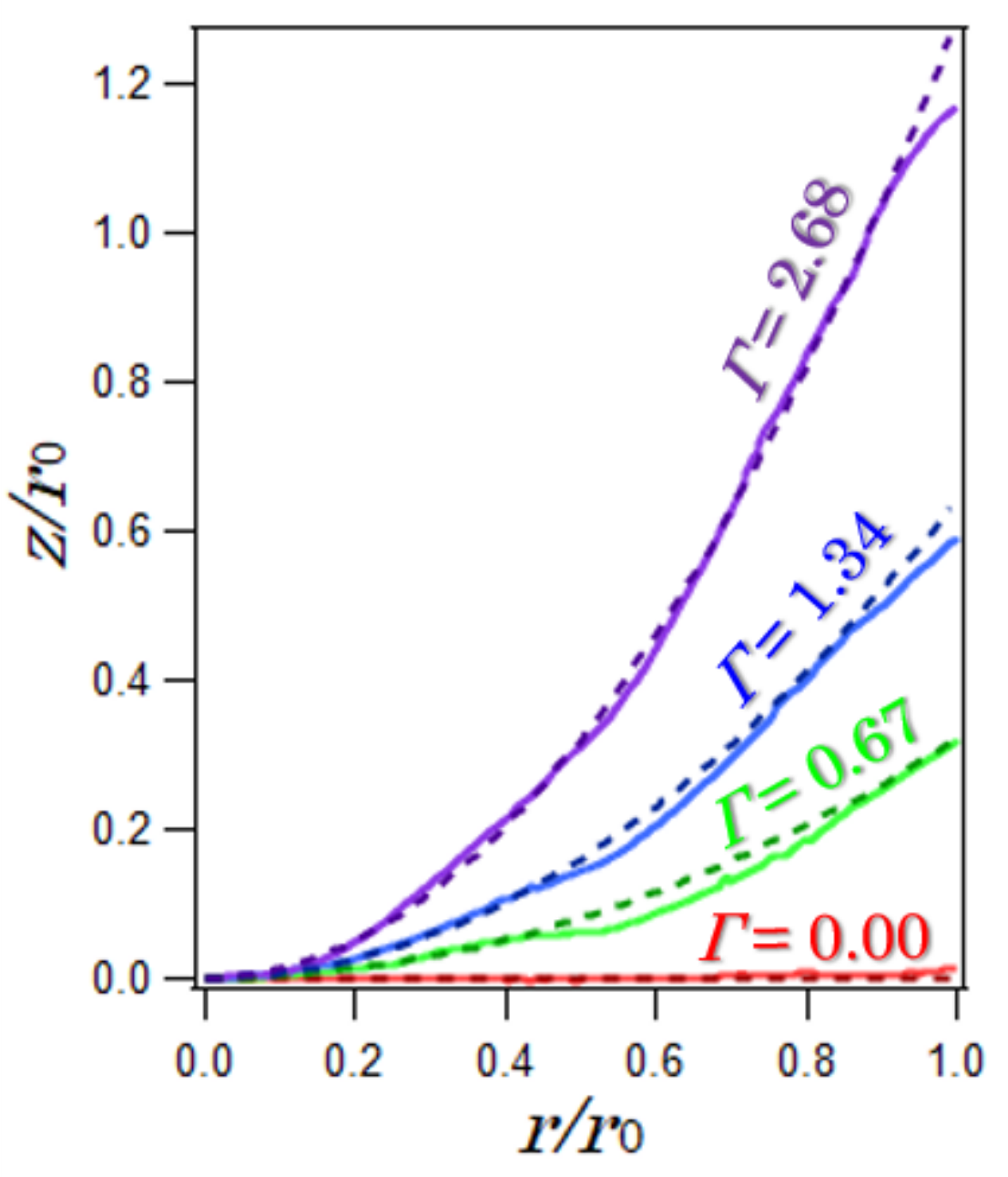}
\end{center}
\caption{Normalized surface profiles of the rotated water layer. Both the height $z$ and distance from the axis $r$ are normalized to the half width of the cell $r_0=50$~mm. $z$ is shifted so that $z(r=0)=0$. The solid curves and dashed curves indicate the averaged experimental profiles and theoretical curves~(Eq.~(\ref{eq:water})), respectively. Very good agreement between experimental results and theoretical curves can be confirmed. Note that this agreement is achieved without any fitting parameter.}
\label{fig:water_prof}
\end{figure}

Perimeters of the water surface can be extracted from the raw images by using ImageJ software. The obtained surface profiles from the center ($r/r_0=0$) to the side edge of the cell ($r/r_0=1$) are shown in Fig.~\ref{fig:water_prof}. By assuming the mirror symmetry, the averaged profiles are shown in Fig.~\ref{fig:water_prof}. The vertical levels of the profiles are adjusted to be $z(r/r_0=0)=0$, i.e., $z_c=0$ for all profiles. One can confirm the reasonable agreement between the experimental data (solid curves) and the model curves of Eq.~(\ref{eq:water}) (dashed curves). Note that the model curves do not have any fitting parameter. All the parameter values are determined by the experimental conditions. Therefore, the excellent agreement strongly suggests the consistency between the experimental result and the theoretical model. The theoretical model of Eq.~(\ref{eq:water}) implies that the local horizon of the water surface (tangent of the surface profile) is normal to the resultant force of gravitational and centrifugal forces. By this experimental apparatus, we can vary the direction of the resultant force by both effects of $r$ (position) and $\omega$ (rotation). Using this feature, deformation by various body-force conditions can be investigated with this system.

By performing the above-mentioned calibrations and test measurements, we are confident with the appropriate measurement of the surface deformed by the centrifugal force. Using the developed system, we measure the deformation of granular samples in the following.

\section{Experimental results}
\subsection{Deformation of glass beads}
\label{sec:glass_beads}
\begin{figure}
\begin{center}
\includegraphics[width=1.0\linewidth]{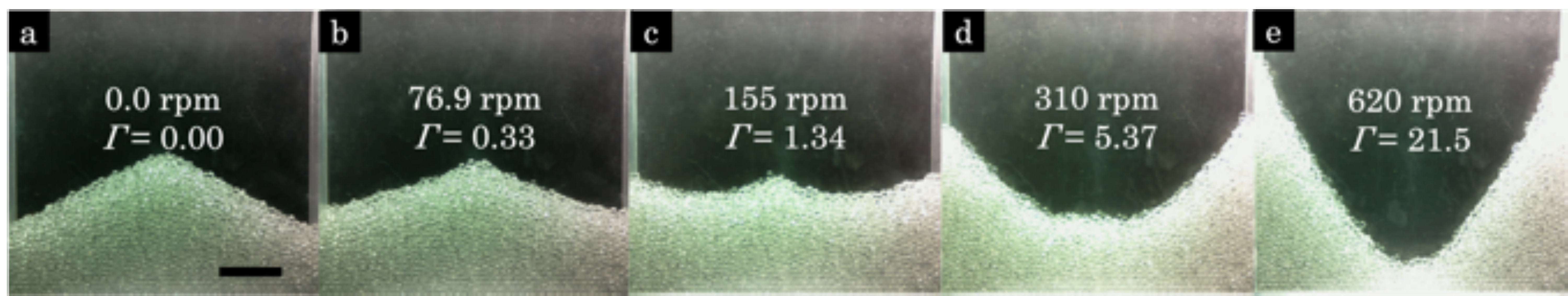}
\end{center}
\caption{Relaxation of the rotated glass-beads pile. Experimental conditions (rotation rates and $\Gamma$ values) are written in each panel. The scale bar indicates 20~mm. As $\Gamma$ increases, the gradual deformation of the pile proceeds. When $\Gamma$ exceeds unity, the granular slopes start to develop on both side walls.}
\label{fig:gb_raw}
\end{figure}

As a first example, we use glass beads of diameter $d=1.5$--$2.5$~mm (ASONE, BZ-2). This grain diameter is sufficiently large to neglect the electrostatic effect which might stick grains on the wall. We find that 1~mm dense particles can be practically used to neglect electrostatic effect. A pile with angle of repose is prepared as shown in Fig.~\ref{fig:gb_raw}(a). The initial pile is made by dropping glass beads using a funnel. The pile is carefully prepared so that its apex corresponds to the center of the cell (rotational axis). Next, the cell is mounted on the rotation apparatus. When the cell is attached to the apparatus, special care is taken not to disturb the structure of the initial granular pile. The rotation rate is gradually raised to achieve the slow deformation (relaxation) of granular pile. Specifically, the dimensionless centrifuge $\Gamma$ is stepwisely increased as $\Gamma=0$, $0.02$, $0.03$, $0.04$, $0.06$, $0.08$, $0.1$, $0.16$, $0.23$, $0.33$, $0.47$, $0.66$, $0.95$, $1.34$, $1.89$, $2.68$, $3.81$, $5.37$, $7.61$, $10.68$, $15.2$, $21.5$. At each step of $\Gamma$, the shape of granular pile is taken by the camera. To attain the equilibrium of surface deformation, rotation rate is slowly increased ($1$~rpm/s) and the image is taken after 14~s from when the inverter's frequency reaches the designated value. We confirm that these conditions are sufficient to quasi-statically equilibrate the pile shape. The obtained raw data of the pile deformation are shown in Fig.~\ref{fig:gb_raw}. We performe three experimental runs to evaluate the reproducibility

As seen in Fig.~\ref{fig:gb_raw}, gradual relaxation of the granular pile can be observed in relatively small $\Gamma$ regime. At $\Gamma \simeq 1$, nearly horizontal surface is formed~(Fig.~\ref{fig:gb_raw}(c)). However, the apex of the initial pile remains as a cusp because of the small centrifugal force at the center. As $\Gamma$ increases, granular slopes are developed on the side walls. And the initial apex is finally obliterated at $\Gamma \simeq 5$~(Fig.~\ref{fig:gb_raw}(d)). Then, the granular surface shape becomes more or less similar to water surface case~(Fig.~\ref{fig:gb_raw}(e)). This gradual deformation of the rotated granular pile shows non-trivial manner of the pile relaxation. By examining the force balance including granular friction effect, we can consider a simple model for the surface shape. However, the details of the model and its physical meaning will be reported elsewhere. Here, in this study, we focus on the instrumental methodology and fundamental characterization of the obtained results.  

\begin{figure}
\begin{center}
\includegraphics[width=0.5\linewidth]{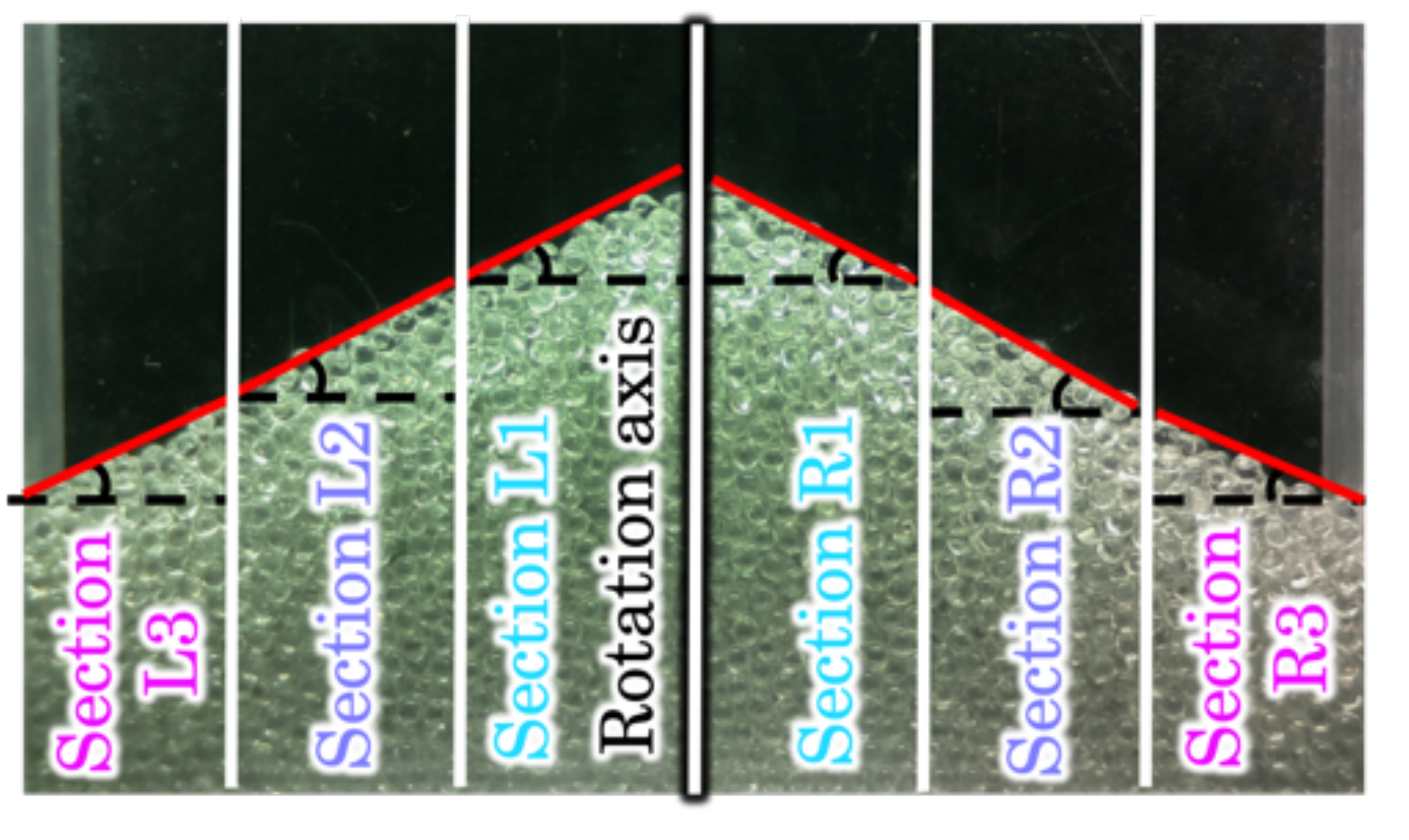}
\end{center}
\caption{Schematic of the horizontal section dividing. The raw image of the granular pile is divided into six sections around the rotation axis. The width of each section approximately corresponds to $17d$. This width is determined to stably estimate the local slopes. The local slopes are calculated by the linear fittings. The positive angle direction is mirror-symmetrically defined. }
\label{fig:section}
\end{figure}

To consider the physics of pile relaxation, we measure the local slopes. The similar analysis has been applied to the analysis of the vibration-induced granular-pile relaxation~\cite{Tsuji:2018}. As shown in Fig.~\ref{fig:section}, the granular pile is horizontally divided into 6 sections. From the center to the edge, sections R1-R3 and L1-L3 are considered in right and left sides of the pile, respectively (Fig.~\ref{fig:section}). The width of each section corresponds to about $17d$. To measure the stable granular slopes (by neglecting grain-scale fluctuation), this width is appropriate. In each section, slope of the granular pile $\theta_i$ ($i=1,2,3$ for left and right) is measured as schematically indicated in Fig.~\ref{fig:section}.  

\begin{figure}
\begin{center}
\includegraphics[width=0.7\linewidth]{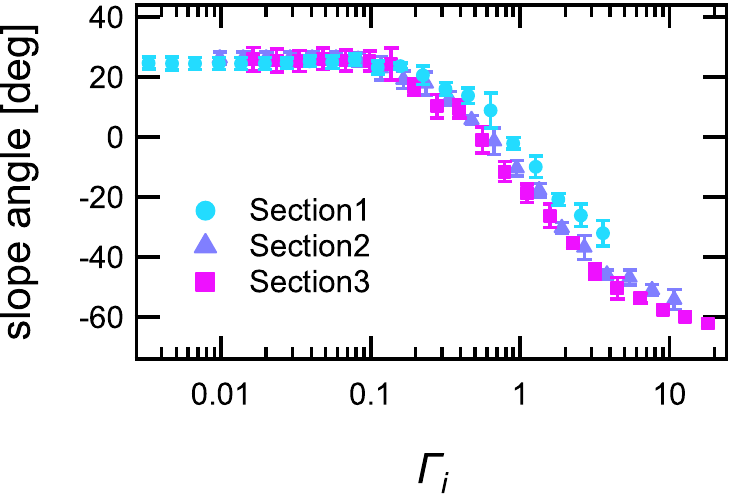}
\end{center}
\caption{Relation between local slope angle $\theta_i$ and local centrifuge degree $\Gamma_i$ computed from the deformation of the glass-beads pile. Errorbars indicate the standard deviation of six profiles (left and right sides of three experimental runs). Basically, all curves seem to obey a simple master relation. At very small $\Gamma_i$ regime, detectable deformation of the slopes cannot be observed. By identifying the beginning of deformation ($\Gamma\simeq 0.1$), the effective cohesion strength can be estimated.}
\label{fig:gb_slopes}
\end{figure}

The measured results of $\theta_i$ for various rotation rates are shown in Fig.~\ref{fig:gb_slopes}. Average slopes of three experimental runs are shown in in Fig.~\ref{fig:gb_slopes}. The abscissa of Fig.~\ref{fig:gb_slopes} indicates the local strength of centrifugal force, $\Gamma_i = r_i \omega^2/g$, where $r_i$ is $r$ at the center in each section, i.e., $r_1=8.3$~mm, $r_2=25$~mm, and $r_3=42$~mm. As can be seen in Fig.~\ref{fig:gb_slopes}, $\theta_i(\Gamma_i)$ seems to obey a master curve which is independent of the distance from the rotation axis. This tendency suggests that the granular-pile relaxation is determined only by the local force balance. In large $\Gamma_i$ regime, the complementary angle ($90^{\circ} + \theta_i$) approaches to the angle of repose (initial $\theta_i$). This means that the granular-pile deformation is mainly governed by the centrifugal force in this regime. In other words, the slope with angle of repose is developed on the side walls due to the large centrifugal force ($r_0\omega^2 \gg g$). This simple behavior could originate from the nature of relatively large glass beads: they are spherical and non-cohesive.

\subsection{Deformation of cohesive dust particles}
\label{sec:dust}
To check the universality of the tendency observed in glass beads (Fig.~\ref{fig:gb_slopes}), we perform the same experiment using cohesive grains. The simplest way to increase the effect of cohesion in granular experiments is using small particles (e.g., \cite{Kleinhans:2011}). However, in the current experimental apparatus, it is difficult to directly use the small particles because small particles are easily stuck on the acrylic front plate. Therefore, we use hierarchical granular matter as a cohesive sample. The hierarchical granular matter consists of $d \simeq 1.7$~mm (sifted by sieving in $1.4 < d <2.0$~mm) aggregates of small ($5~\mu$m in diameter) glass beads~\cite{Vazquez:2020}. Each macroscopic aggregate particle has about 70\% porosity. By assuming typical random packing porosity $\simeq$40\%, very porous structure with bulk porosity $\simeq$82\% is achieved for hierarchical granular matter. Although the hierarchical granular matter is cohesive, its cohesive effect is not very strong~\cite{Katsuragi:2018}. Thus, this type of dust particles (hierarchical granular matter) could be applicable to the current setup. Moreover, such a collection of porous dust aggregates could play a key role for understanding the terrain dynamics on small bodies in the solar system~\cite{Skorov:2012,Bentley:2016,Okada:2020}. Therefore, we employ dust particles as representative cohesive test materials.

\begin{figure}
\begin{center}
\includegraphics[width=1.0\linewidth]{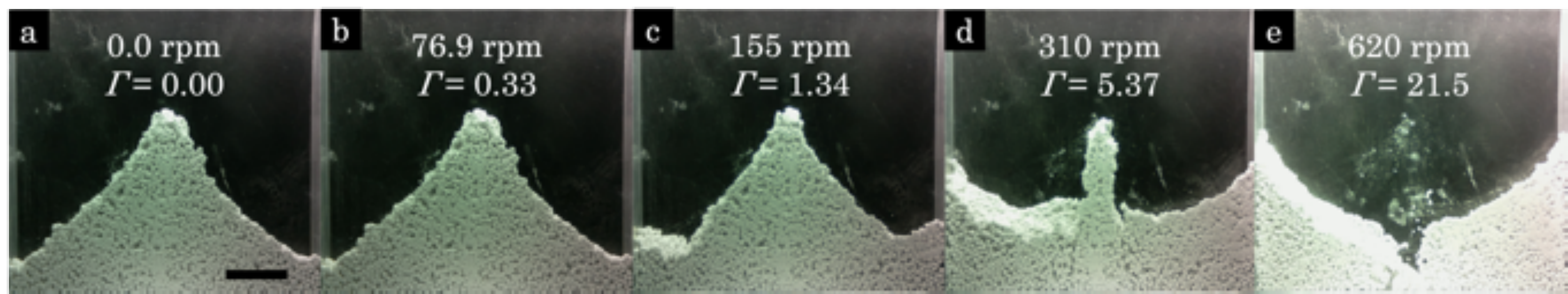}
\end{center}
\caption{Relaxation of the rotated dust-particles pile. Experimental conditions (rotation rates and $\Gamma$ values) are written in each panel. The scale bar indicates 20~mm. The deformation process is quite different from the glass-beads case shown in Fig.~\ref{fig:gb_raw}. A vertical pillar structure is retained at the center in relatively small $\Gamma$ regime~(c,d). At large $\Gamma$ regime, fracturing of the central part is observed. These behaviors arise from the cohesion effect and its inhomogeneity in the dust-particles pile.}
\label{fig:dust_raw}
\end{figure}

The experimental result obtained by dust particles (hierarchical granular matter) is shown in Fig.~\ref{fig:dust_raw}. One can clearly confirm that the deformation process shown in Fig.~\ref{fig:dust_raw} is quite different from that of glass beads (Fig.~\ref{fig:gb_raw}). While the deformation of glass-beads pile is gradual and smooth, dust-particles pile shows strongly inhomogeneous deformation. This complexity directly comes from the small density and inhomogeneous cohesion effect. On the rotated glass-beads pile, small avalanching flows gradually modify the shape of the pile. However, on the rotated dust-particles pile, a cluster of particles is suddenly detached (rather than flowing) at a certain place. By repeating this process, the global structure of dust-particles pile is deformed. Moreover, although the slopes on the side walls are finally developed in large $\Gamma$ regime, their shapes are quite different from parabola (Fig.~\ref{fig:dust_raw}(e)). Besides, it also exhibits the fracturing around the rotation axis. Even in the intermediate $\Gamma$ regime, a vertical pillar structure is formed at the center (Fig.~\ref{fig:dust_raw}(d)). This pillar structure is also a consequence of cohesion effect. In addition, attrited small particles are stuck on the front wall presumably due to the electrostatic force (Fig.~\ref{fig:dust_raw}(d,e)). 

\begin{figure}
\begin{center}
\includegraphics[width=0.7\linewidth]{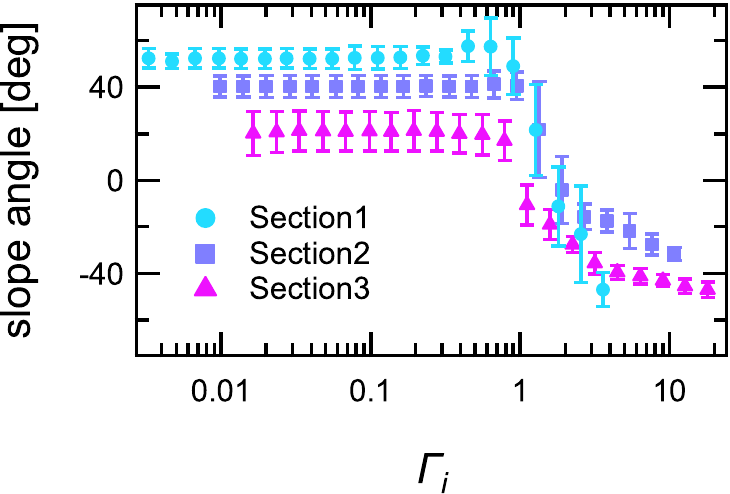}
\end{center}
\caption{The relation between local slope angle $\theta_i$ and local centrifuge degree $\Gamma_i$ computed from the deformation of the dust-particles pile. Errorbars indicate the standard deviation of six profiles (left and right sides of three experimental runs). Compared with the glass-beads case (Fig.~\ref{fig:gb_slopes}), the data scattering due to the heterogeneous cohesion effect is significant. The slope relaxation is triggered by the larger $\Gamma$ value ($\Gamma\simeq 1$) compared with the glass-beads case. }
\label{fig:dust_slopes}
\end{figure}

In order to quantitatively assess the difference between the rotated glass beads and dust particles, temporal development of the local slopes for dust particles are also measured. The average results obtained by three experimental runs are shown in Fig.~\ref{fig:dust_slopes}. The global trend seen in Fig.~\ref{fig:dust_slopes} is slightly similar to that shown in Fig.~\ref{fig:gb_slopes}. However, the position ($i$) dependence of $\theta_i(\Gamma_i)$ can be confirmed rather than the data collapsing to the master curve. In addition, the slope relaxation is triggered at $\Gamma_i \simeq 1$ (Fig.~\ref{fig:dust_slopes}). This behavior is contrastive to the glass-beads case (Fig.~\ref{fig:gb_slopes}) in which the gradual relaxation of the slope can be confirmed even in $\Gamma_i < 1$. Note that, however, a small but non-negligible $\Gamma$ ($\Gamma_i \simeq 0.1$) is required to induce the relaxation of the initial pile even in the glass-beads case. Using this feature, we can roughly estimate the cohesion strength of granular matter. 

\section{Discussion}
\label{sec:discussion}
By assuming that the cohesion effect stabilizes the initial pile, we can estimate the effective cohesion strength of the granular pile. Moreover, this strength could be a crucial quantity characterizing the difference between starting angle (static angle) and angle of repose (dynamic angle) of granular slopes. In other words, we consider that the effective cohesion developed in a settled granular pile could stabilize its structure until the slope becomes the starting angle. Using the current experimental result, we can quantify this effect by considering the competition between gravity and centrifuge. Let us consider a granular slope with angle of repose $\theta_\mathrm{r}$ at position $i$. Obviously, this slope is stable without rotation. By increasing the rotation rate, the surface deformation is triggered at a certain rotation rate. We assume that the cohesive force balances with the centrifuge-originated slipping force at this instance. Then, from the force balance per unit mass along the slope, we obtain a relation, 
\begin{equation}
  r_i\omega^2 \cos\theta_r + g \sin \theta_\mathrm{r} = \mu(g \cos \theta_\mathrm{r} - r_i \omega^2 \sin\theta_\mathrm{r} ) + a_\mathrm{c},
  \label{eq:force_balance}
\end{equation}
where $\mu=\tan \theta_\mathrm{r}$ and $a_\mathrm{c}$ are the bulk friction coefficient defined by angle of repose and the effective acceleration by cohesive force per unit mass, respectively. By this definition, effective shear strength $C$ can be estimated as $C=\rho a_\mathrm{c} d$, where $\rho$ is the bulk density of granular matter.  Here we assume a single particle layer (thickness $d$) is moved by the effect of centrifuge at the beginning of macroscopic deformation. By using the above notations, we finally obtain a simple relation,
\begin{equation}
C = \rho g d \Gamma_i \sqrt{1 + \mu^2} = \rho d r_i \omega^2 \sqrt{1+\mu^2},
\label{eq:strength}
\end{equation}
To calculate $C$, we use $\rho=1500$~kg/m$^3$, $\mu(=\tan \theta_r)=0.47$, and $d=2.1$~mm for glass beads and $\rho=630$~kg/m$^3$, $\mu=1.0$, and $d=1.7$~mm for dust particles. These values are directly measured in the experiment. From Figs.~\ref{fig:gb_slopes} and \ref{fig:dust_slopes}, the $\Gamma_i$ value at which the macroscopic deformation is triggered can be identified in each section. Specifically, sudden slope change with more than 3$^{\circ}$ is defined as the beginning of macroscopic deformation (because the slope fluctuation level is about 2$^{\circ}$).  Using these quantities, we obtain $C=6 \pm 2$~Pa and $18 \pm 2$~Pa for glass beads and dust particles, respectively. Although the difference in $C$ is not very significant, qualitative deformation manners of glass-beads pile (Fig.~\ref{fig:gb_raw}) and dust-particles pile (Fig.~\ref{fig:dust_raw}) are quite different. Namely, the profile deformation could strongly depend on the cohesiveness and its inhomogeneity. 

Effective cohesion strength can also be estimated from the central pillar structure shown in Fig.~\ref{fig:dust_raw}(d). From the simple force balance between cohesion and centrifugal force, a relation $C'=\rho (r' \omega)^2$ is obtained, where $r'=5.6$~mm is the measured radius of the central pillar. Substituting $\rho=630$~kg/m$^3$ and $\omega= 32.6$~rad/s~$=310$~rpm~(Fig.~\ref{fig:dust_raw}(d)), one can obtain another effective cohesion strength $C' = 21$~Pa. This value is indeed close to $C=18$~Pa. Because $C=6$~Pa for glass beads is about 1/3 of that for dust particles, the expected pillar width in glass-beads pile at $\omega=32.6$~rad/s becomes $r'=1.9$~mm. This value is comparable to the particle diameter. Thus, pillar-like structure cannot be observed in Fig.~\ref{fig:gb_raw}(d). Actually, the obtained range of cohesion strength ($10^0$--$10^1$~Pa) is more or less close to the strength of surface materials on the surface of asteroid Ryugu that was estimated from the size of artificial impact crater~\cite{Arakawa:2020}. 

More directly, the centrifugal force applied to a dust particle of $d=1.7$~mm at $r'=5.6$~mm with $\omega=32.6$~rad/s and $\rho=630$~kg/m$^3$ can be estimated as $F=(\pi/6)d^3\rho r'\omega^2= 9.6 \times 10^{-6}~N$. 
Recently, a simple method to estimate the number of monomer contacts between connecting dust aggregates was proposed~\cite{Arakawa:2020s}. By assuming glass surface energy $25$~mN/m and monomer radius $2.5$~$\mu$m, single contact force of monomers used in this study is estimated as $F_\mathrm{single}=3 \times 10^{-7}$~N. Then, the number of contacts is estimated as $F/F_\mathrm{single} =32$. Namely, multiple contacts result in the cohesion strength in dust particles. According to Ref.~\cite{Nagaashi:2021}, the actual cohesion force under the atmospheric condition might become about one order of magnitude below the theoretical estimate. Thus, several hundreds contact may exist in actual dust-particles layer.

Using a dimensionless number $R_g=\rho g d/C$, we can compare the effects of gravity and cohesion strength. When $R_g \gg 1$, gravity dominates the dynamics while the strength becomes important in $R_g \ll 1$. From the values estimated above, we obtain $R_g=5.1$ and $0.58$ for glass beads and dust particles, respectively. While both values are not significantly greater (or less) than unity, the former (glass beads) indicates the gravity-dominant regime while the latter (dust particles) corresponds to the strength-dominant regime. This difference causes the qualitative difference between Figs.~\ref{fig:gb_raw} and \ref{fig:dust_raw}. 

By the effect of centrifugal force, direction of local gravity can be tilted. From the geometrical condition, the tilting angle $\theta_g$ can be estimated by $\theta_g = \arctan(r_i \omega^2/g)$. By substituting $\Gamma_i=(r_i \omega^2/g) \simeq 0.1$ (glass-beads case), we obtain $\theta_g \simeq 6^{\circ}$. This value is larger than usual difference between starting angle and angle of repose~\cite{Nagel:1992}. Namely, the weak but finite strength larger than the difference of the starting angle and the angle of repose could be measured by this experiment. However, further studies about the stability of granular pile using the centrifugal force are necessary to conclude the relation between two angles and cohesion.

Hysteresis in granular-pile deformation can also be quantitatively confirmed in this study. When we use water for the rotated sample (Sec.~\ref{sec:water_deform}), surface parabola shape is determined only by the parameter $\Gamma$. We can confirm the shape is identical as long as the value of $\Gamma$ is same, i.e., $\Gamma$ is regarded as a state variable. This characteristic can be simply checked by repeating the increase and decrease of $\Gamma$. Water surface deformation does not show any history dependence. However, this is not the case when granular matter is used. Due to the nonlinearity of granular friction, history dependence is introduced in a rotated granular-pile deformation. Obviously, the deformation is irreversible. When we turn off the rotation after $\Gamma=21.5$, the initial pile configuration (the initial state of $\Gamma=0$) cannot be recovered. Moreover, the stability of the granular pile could be affected by the loading of the centrifuge. Such a protocol dependence of the rotated granular pile is a future problem to be considered. 
 
Although we neglect three-dimensional effect, it slightly affects the experimental result. As the rotation rate increases, the surface particles begin to move. Then, the particles move to the side wall due to the centrifugal force. At the same time, particles move to the depth direction (perpendicular to both $r$ and $z$ direction) due to the inertial effect. Then, the slope is developed also in the thickness direction. This effect is particularly visible as bright parts in the left half of the dust-particles case (Fig.~\ref{fig:dust_raw}). The same effect causes the slight surface tilting in the glass-beads case, too. However, the tilting is almost uniform on the surface of the glass-beads pile. Thus, we neglect its effect as a first-tep approximation. Whereas a much thinner cell could eliminate this effect, the effect of wall friction is alternatively enhanced in the thinner cell. In this study, we use a relatively thick cell to reduce the wall friction effect. Put differently, the effect of wall friction does not play a significant role in this experiment, at least less significant than the inertial effect.  

In a granular flow driven by continuously supplied particles, only a thin surface layer is fluidized~\cite{Lemieux:2000,Komatsu:2001,Katsuragi:2010}. However, vibro-fluidized granular-pile relaxation results in thicker fluidization layer (approximately the upper half of the pile is fluidized)~\cite{Tsuji:2019}. Measurement of the thickness of fluidized layer is crucial to characterize the deformation of granular layer. Characterization of the internal flow in the rotated granular pile is another important future problem.

In this paper, only two kinds of particles (glass beads and dust particles) are used to compare the effect of cohesion. The effects of particle shape and density on the deformation of rotated granular pile are not discussed. These effects are definitely important to discuss natural phenomena such as asteroidal top shape etc.

\section{Conclusion}
In this study, an experimental apparatus by which granular surface deformation due to the centrifugal force can be measured was developed. The apparatus rotates the quasi two-dimensional cell around the vertical axis. After assembling the apparatus, image calibration and test measurement using water were carefully performed to establish the reliable methodology. Using the developed setup, deformation of granular piles made of glass beads or dust particles were observed. Due to the position-dependent centrifugal force, the resultant granular-surface shapes show nontrivial curves. By analyzing the local slopes, however, the unified relation between local slope and local centrifugal effect was confirmed for the glass-beads case. For dust-particles case, local slopes show considerable scattering owing to the strong cohesive effect and its inhomogeneity. By introducing the threshold of surface deformation, the effective cohesion strength of granular matter was estimated. While there are some limitations and future problems to be solved, the newly developed system has a great potential to characterize granular mechanics.

\ack
SW and HK thank JSPS KAKENHI Grants No.~19H01951 and HK thanks JSPS KAKENHI Grants No.~18H03679 for financial support. SW also thanks JSPS KAKENHI Grants No.~17H06459.

\section*{References}
\bibliography{rot_relax}

\end{document}